\documentclass[epj-spec]{svjour}
\usepackage{graphicx}
\usepackage{times}
\usepackage{bbm}
\newcommand{\sinc}{{\rm sinc}}

\begin{document}
\title{Micromaser line broadening without photon exchange}
\author{Carsten Henkel%
%\inst{1}
\fnmsep\thanks{\email{henkel@uni-potsdam.de}}}

\institute{Institut f\"{u}r Physik, Universit\"{a}t Potsdam,
Germany}
\abstract{
We perform a calculation of the linewidth of a micromaser, using 
the master equation
and the quantum regression approach. A `dephasing' contribution is 
identified from pumping processes that conserve the photon number 
and do not appear in the photon statistics. We work out examples for 
a single-atom maser with a precisely controlled coupling and for a 
laser where the interaction time is broadly distributed. 
In the latter case, we also assess the convergence of a 
recently developed uniform Lindblad approximation to the master 
equation; it is relatively slow.
} %end of abstract
\maketitle
\section{Introduction}
\label{intro}

The linewidth of a laser and the spectrum of a resonantly 
driven two-level system are paradigmatic
examples that probably spawned the field of quantum 
optics~\cite{Walther98c,Loudon,Orszag}. In both cases, one 
has to deal with an open quantum system that is driven towards a 
steady state by the balance between pumping and dissipation.
The characteristic features are intimately linked to the 
quantized nature of the electromagnetic field, with spontaneous 
processes determining the (laser) linewidth and the triplet spectrum 
(resonance fluorescence). 

A fertile playground to investigate these issues in great detail
is the micromaser or single-atom laser~\cite{Meschede85,Brune87}.
Its basic ingredients are 
well known: a high-$Q$ cavity sustaining an electromagnetic field 
mode, and a dilute atomic beam that delivers energy by entering the cavity 
in an electronically excited state. In this device, a delicate 
control of the interaction time is possible and permits the 
observation of non-classical behaviour of the cavity field like 
quantum jumps~\cite{Benson94a} or the collapse into photon number 
states~\cite{Weidinger99,Varcoe00}. In addition, the atoms leaving the 
cavity provide information that can be used, for example, to measure 
in a non-destructive way the cavity photon 
number~\cite{Turchette95,Nogues99}.

%\cite{Nogues99} %Seeing a single photon without destroying it
%\cite{Benson94a} %Quantum jumps of the micromaser field: Dynamic behavior 
                 %close to phase transition points
%\cite{Weidinger99} %Trapping states in the micromaser
%\cite{Varcoe00} %Preparing pure photon number states of the radiation field
%\cite{Brattke01} %Generation of Photon Number States on Demand via Cavity 
                 %Quantum Electrodynamics
%\cite{Casagrande03} %How to Measure the Phase Diffusion Dynamics in the Micromaser
%\cite{Turchette95} %Measurement of Conditional Phase Shifts for Quantum Logic

In this paper, we discuss the linewidth of the micromaser or in other 
words, the coherence time of the cavity field. This has been studied a 
few years after the introduction of the basic principles of the 
micromaser~\cite{Filipowicz86} by Scully and 
co-workers~\cite{Scully91}. The basic ideas are simple: the maser (or 
laser) spectrum is defined by the Fourier transform of the two-time 
correlation (or coherence) function of the cavity field 
operator~\cite{Scully68}. To calculate 
these correlations, one invokes the so-called `quantum regression 
formula'~\cite{Lax63,Lax00} that provides a link to the time evolution of 
the system density operator. This evolution is conveniently given in 
terms of a master equation whose eigenstates and eigenvalues determine
the maser spectrum. Their computation typically requires 
numerical work, and it is difficult to find a physical interpretation 
of the results. In particular, more than one eigenvalue contributes 
to the result around the trapping states where 
phase transitions happen in the micromaser~\cite{Vogel93}. 
We suggest here a formulation of the maser linewidth (i) that provides a 
simple physical interpretation; (ii) that can be computed fairly easily 
once the steady state solution to the master equation is known; and
(iii) that interpolates in a natural way between different 
eigenvalues of the master equation. Our 
methods are similar to those used by Scully~\cite{Scully91} and McGowan
and Schieve~\cite{McGowan97} where a `linewidth operator' that depends 
on the photon number is identified, whose expectation value with 
respect to the steady-state density operator yields the linewidth. The 
main difference of our approach is its straightforward formulation and 
simplicity. The results are generally in good agreement with previous 
work. We predict a weak oscillation of the linewidth as a function of 
the pumping strength that is related to a dephasing process related to 
the non-destructive monitoring of the cavity photon number by atoms 
that do not deposit their excitation energy in the cavity.

The outline is as follows. We fix the notation in Sec.\ref{sec:1}, 
introduce our formula for the linewidth (Eq.(\ref{eq:read-off-linewidth}) 
of Sec.\ref{sec:def-spectrum})
and work out the details for two cases: a `perfect maser' where the 
interaction time $\tau$ is fixed (Sec.\ref{s:maser}) and a
`conventional laser' where $\tau$ is exponentially distributed
(Sec.\ref{s:laser}). We conclude by a comparison to a recently 
developed approximation~\cite{Henkel07a} that turns out to describe 
the laser above threshold fairly well, but becomes increasingly 
complex in the strong pumping regime (Sec.\ref{s:uniform}). The 
Appendix contains some technical details of the calculation of the 
micromaser linewidth.

\section{The micromaser and its linewidth}

\subsection{Master equation}
\label{sec:1}

The master equation for the micromaser we use here is of the form
\cite{Lugiato87,Orszag}
\begin{eqnarray}
    \frac{ {\rm d}\rho }{ {\rm d}t } &=&
    \kappa ( 1 + n_{\rm th})\left[
    a \rho(t) a^\dag - {\textstyle\frac{ 1 }{ 2 }}
    \left\{ \rho(t) a^\dag a +  a^\dag a \rho(t) \right\}
    \right]
    \label{eq:master-eqn}\\
    && {} +
    \kappa n_{\rm th} \left[
    a^\dag \rho(t) a - {\textstyle\frac{ 1 }{ 2 }}
    \left\{ \rho(t) a a^\dag +  a a^\dag \rho(t) \right\}
    \right]
    \nonumber\\
    && {} +
    \int\!{\rm d}p( \tau )
    \left[
    r \cos( g \tau \hat \varphi ) \rho( t ) \cos( g \tau \hat \varphi ) 
    - r \rho( t )
    + \kappa \theta^2
    a^\dag \sinc( g \tau \hat \varphi ) \rho( t ) 
    \sinc( g \tau \hat \varphi ) a
    \right]
    \nonumber
\end{eqnarray}
The first two lines describe cavity damping and its thermal excitation
by the surrounding in the usual way. The thermal occupation number 
$n_{\rm th}$ depends on temperature and gives the average photon 
number at equilibrium, without pumping.
%$n_{\rm th} = ({\rm e}^{ \hbar \omega_{\rm c} / k_{\rm B} T} - 1)^{-1}$
The maser
is pumped by a stream of excited two-level atoms that enter the cavity
one by one with Poissonian statistics (rate $r$).  The atoms interact 
with the maser mode for a time $\tau$
that is distributed according to the measure ${\rm d}p( \tau )$, and
$g$ is the one-photon Rabi frequency for this interaction.  The
operator $\hat\varphi = (a a^\dag)^{1/2}$ and $\sinc(x) = \sin(x)/x$.
The conventional pumping parameter is given by
$\theta = (r/\kappa)^{1/2} g \tau$. 

In the strong coupling regime, non-classical states of the radiation
field like number states are generated whenever the photon number 
$n$ satisfies $g \tau \sqrt{ n + 1 } = \pi,
2\pi, \ldots$ \cite{Filipowicz86}.
Then, the last term in Eq.(\ref{eq:master-eqn}) is
zero and the photon number in the cavity remains unchanged.
Physically speaking, each incoming atom performs an integer number of
full Rabi cycles and leaves the cavity in the excited state again.
This perfect scenario is perturbed by the fluctuations of the 
interaction time $\tau$ and a not perfectly controlled coupling 
strength $g$ that depends, for example, on the atom crossing the laser 
mode at a node or an anti-node.  A typical choice for the distribution
${\rm d}p( \tau )$ in conventional lasers is an exponential one, its 
mean value $\bar\tau$ giving the lifetime of the excited atomic 
states. 

% In the following, 
% particular form for ${\rm d}p( \tau )$? peaked?

\subsection{Maser spectrum}
\label{sec:def-spectrum}

The maser spectrum is defined by the Fourier transform 
of the two-time correlation function~\cite{Orszag,Scully68}
\begin{equation}
    g( t ) = \langle a^\dag( t_0 + t ) a( t_0 ) \rangle 
    \label{eq:def-two-time-correlation}
\end{equation}
and is usually computed in the stationary regime 
where the correlation function only depends on the time difference
$t$ with respect to which the Fourier transform is taken. We 
therefore drop the $t_0$ argument from Eq.(\ref{eq:def-two-time-correlation}).
% 
% Spectrum of the laser mode: Fourier transform with respect to $t$, in 
% the limit $T \to \infty$ (stationary regime). 
Alternative, but equivalent definitions are based on the decay of the
average electric field operator $\langle a( t )\rangle$
(Ref.\cite{Scully91}) or the correlation function of ${\rm e}^{ {\rm
i} \phi( t )}$ with $\phi$ being the phase operator
(Ref.\cite{Quang93}).

Using arguments in the spirit of the quantum regression formula,
one shows that the correlation function can be computed as the
expectation value \cite{Scully68,Quang93,Vogel93}
\begin{equation}
    g(t ) = {\rm tr}\left[ a^\dag P( t ) \right]
    \label{eq:q-regression-1}
\end{equation}
of a photon creation operator with respect to a `skew density
operator' $P$. The latter obeys an equation of motion identical 
to the master equation~(\ref{eq:master-eqn}), but  with the 
`initial condition' $P(0) = a \rho_{\rm ss}$ where $\rho_{\rm ss}$ is 
the stationary solution that turns out to be diagonal in the photon 
number basis. A trivial consequence is the normalization to the 
average photon number $g( 0 ) = 
{\rm tr}\left( \hat n \rho_{\rm ss} \right) \equiv
\langle \hat n \rangle_{\rm ss}$.

The expansion of $P( t )$ in eigenvectors of the master equation 
yields a sum of exponentials in time and hence of Lorentzian spectra. 
Usually, the spectra are computed in the frequency domain and their 
characteristic width computed numerically. 
We adopt here a different approach that gives straightforward 
analytical results, but restricted to the regime where at most 
two eigenvalues are dominating.  We emphasize this happens over a
relatively large part of the parameter space (see Ref.\cite{Vogel93}).
Our definition of the maser linewidth $D$ is
\begin{equation}
    - \frac{ D }{ 2 } = \frac{ 1 }{ \langle \hat n \rangle_{\rm ss} }
    \left. \frac{ {\rm d}g }{ {\rm d}t } \right|_{0}
    \label{eq:read-off-linewidth}
\end{equation}
where the time derivative is computed directly from the master
equation, using the quantum regression formula~(\ref{eq:q-regression-1}).
To justify this choice, note that the eigenvector expansion for the 
correlation function is of the form
\begin{equation}
    g( t ) = \sum_{j} g_{j} {\rm e}^{ - \mu_{j} t / 2 }
    \label{eq:eigenvalue-expansion}
\end{equation}
where the prefactors satisfy the sum rule $\sum_{j} g_{j} = 
g( 0 ) = \langle \hat n \rangle_{\rm ss}$. This leads to 
\begin{equation}
    D = \frac{ \sum_{j} g_{j} \mu_{j} }{ \sum_{j} g_{j} }
    \label{eq:width-from-eigenvalues}
\end{equation}
hence an arithmetic mean of the exact eigenvalues. Now, it appears 
from Ref.\cite{Vogel93} that the relative weights $g_{j} / 
\langle \hat n \rangle_{\rm ss}$ are often close to either zero or one.
Only when the 
two lowest eigenvalues are crossing, both are 
contributing significantly. If a single eigenvalue is 
relevant, our linewidth $D$ coincides with it; if two contribute, 
Eq.(\ref{eq:width-from-eigenvalues}) is just another convenient way 
to combine the two into a single parameter. We note that the 
conventional full width at half maximum (FWHM) for a sum of two
Lorentzians of equal weights, but of different widths, leads to the
geometric mean of the widths.  This significantly differs from the
arithmetic mean only when the widths are very different.  But, as
mentioned above, equal weights occur typically when the eigenvalues
cross.

\subsection{Explicit formula and comparison}

The time derivative in Eq.(\ref{eq:read-off-linewidth}) leads to,
using the regression formula~(\ref{eq:q-regression-1})
\begin{equation}
D = - \frac{ 2 }{ \langle \hat n \rangle_{\rm ss} }  
{\rm tr}\left[ a^\dag
\left. \frac{ {\rm d} P }{ {\rm d} t } \right|_{0}
\right]
\label{eq:simple-formula-for-linewidth}
\end{equation}
This expression involves the master equation for the skew operator
$P$ which can be calculated easily by adopting a Lindblad form. 
One gets the familiar result
\begin{equation}
    D \langle \hat n \rangle_{\rm ss} = 
    - \sum_{\lambda}
    {\rm tr}
    \left\{
    L_{\lambda}^\dag
    \left[ a^\dag, L_{\lambda} \right] 
    a \rho_{\rm ss} 
    +
    \left[ L_{\lambda}^\dag, a^\dag \right] L_{\lambda}
    a \rho_{\rm ss} 
% 
%     \mbox{(second term)}
%     \left[ a^\dag, L_{\lambda} \right] 
%     a \rho_{\rm ss} L_{\lambda}^\dag
    \right\}
    \label{eq:linewidth-from-Lindblad}
\end{equation}
where the $L_{\lambda}$ are the so-called jump or Lindblad operators.
For the micromaser, they can be written as (the index is running over
$\lambda = 0, 1, ({\rm c},k), ({\rm s},k)$)
\begin{eqnarray}
    L_{0} &=& \sqrt{ \kappa ( 1 + n_{\rm th} )}\, a
    \label{eq:def-Lindblad-start}
    \\
    L_{1} &=& \sqrt{ \kappa \, n_{\rm th}}\, a^\dag
    \\
    L_{{\rm c},k} &=& \sqrt{ r\, \Delta p( \tau_{k} ) }\, 
    \cos( g \tau_{k} \hat \varphi ) + 
    \mathbbm{1} f_{k}
    \\
    L_{{\rm s},k} &=& \sqrt{ r\, \Delta p( \tau_{k} ) }\, 
    g \tau_{k}
    \,a^\dag \sinc( g \tau_{k} \hat \varphi ) 
    \label{eq:def-Lindblad}
\end{eqnarray}
where we have broken the integral over $\tau$ into a 
Riemann sum with weights $\Delta p( \tau_{k} )$.
The term proportional to the unit operator in $L_{{\rm c},k}$ actually
does not contribute to the Lindblad master equation; an expression for
the real number
$f_{k}$ can be found in Ref.\cite{Henkel07a}, it  is not needed here.
% We write $\theta_{k} = (r/\kappa)^{1/2} g \tau_{k}$ for the pumping
% parameter.
(Note that the term $- r \rho$ in Eq.(\ref{eq:master-eqn}) is generated 
by 
the anticommutators in the Lindblad form, after performing the Riemann
integral.)

We immediately see that one gets the linewidth as the average value 
of some operators with respect to the steady-state photon statistics,
similar to what has been found previously~\cite{Scully91,McGowan97}.
In Appendix~\ref{a:work-out-linewidth}, these operators are computed 
and simplified using the detailed balance condition for the steady state.
The result is %[check $\sin$ vs $\sinc$ in second term]
\begin{eqnarray}
    D \langle \hat n \rangle_{\rm ss} &=& 
    \kappa n_{\rm th} + 
    \sum_{k}
    \Delta p( \tau_{k} ) {\rm tr} \left\{
    r \big( \cos(g \tau_{k} \hat \varphi ) -
    \cos(g \tau_{k} \hat n^{1/2} ) \big)^2 \hat n \rho_{\rm ss}
    \right.
    \nonumber\\
    && \left. {} + 
    r \big( \hat \varphi \sin(g \tau_{k} \hat \varphi ) -
    \hat n^{1/2} \sin(g \tau_{k} \hat n^{1/2} ) \big)^2 \rho_{\rm ss}
    \right\}
    \label{eq:result-linewidth}
\end{eqnarray}
where $\hat n = a^\dag a$ is the photon number operator and we have kept
for the ease of comparison the discrete summation. 
% Note that the 
% formulation with 
% result is valid only if $\rho_{\rm ss}$ is diagonal in the Fock basis.

This result should be compared to previous calculations. In their seminal 
paper~\cite{Scully91}, Scully and co-workers make the approximation of
a sufficiently narrow distribution of photon numbers and get a
linewidth
\begin{equation}
    D \approx \frac{ 1 + 2 n_{\rm th} }{ 4 \langle \hat n \rangle_{\rm ss} }
    +
    4 r \sin^2\left( \frac{ g \tau }{ 4 \sqrt{ \langle \hat n 
    \rangle_{\rm ss} } } \right). 
    \label{eq:Scully-linewidth}
\end{equation}
McGowan and Schieve improve this calculation by allowing for multiple
eigenvalues of the Liouville operator~\cite{McGowan97}.  They find an
expression that gives a good approximation (to a few percent) to the
exact eigenvalue calculation.  The method is using a numerical
evaluation of the FWHM of the (non-Lorentzian) spectrum which we want
to avoid here.  But their approximation can also be formulated in
terms of the correlation function $g(t)$, and applying our
definition~(\ref{eq:read-off-linewidth}), one has
\begin{eqnarray}
    D \langle \hat n \rangle_{\rm ss} &\approx& 2
    \sum_{k} \Delta p( \tau_{k} )
    {\rm tr}\left[ \big( r - 
    r  \sin(g \tau_k \hat \varphi) \sin(g \tau_k \hat n^{1/2})
    - r \cos(g \tau_k \hat \varphi) \cos(g \tau_k \hat n^{1/2}) 
    \right.
    \nonumber\\
    && \left. {} +
    \kappa (1 + n_{\rm th}) (\hat n - {\textstyle\frac12} - [\hat n(\hat n-1)]^{1/2} )
    +
    \kappa n_{\rm th} (\hat n + {\textstyle\frac12} - \hat \varphi \hat n^{1/2} )
    \big) \hat n \rho_{\rm ss} \right]
    \label{eq:McGowan-linewidth}
\end{eqnarray}
where we have added the average over $\tau$ as before.

\section{Discussion}
\label{s:discussion}

\subsection{Perfect maser}
\label{s:maser}

\begin{figure}[hbt]
    \centerline{\includegraphics*[width=70mm]{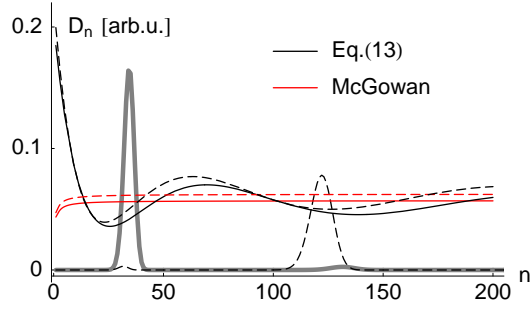}}
    \caption[]{Contribution to the maser linewidth for a given 
    photon number. We plot the $n$-dependent term under the trace
    of Eq.(\ref{eq:result-linewidth}) (undulating curve)
    and Eq.(\ref{eq:McGowan-linewidth}) (`McGowan') . The normalization 
    is arbitrary, and the interaction time is fixed (no average).
    The peaked curves give the steady-state photon statistics.
    Solid and dashed curves correspond to two pumping parameters:
    $\theta = 2.1\,\pi$ ($g \tau \approx 0.47$,
    left peak, solid line) and 
    $\theta = 2.2\,\pi$ ($g \tau \approx 0.49$, right peak, dashed 
    line).  We fix the pumping rate to $r = 200\,\kappa$ and $n_{\rm
    th} = 0.1$.}
    \label{fig:resolved-dephasing}
\end{figure}

To get a better understanding of the physics behind the maser 
linewidth, we first fix the interaction time (no fluctuations)
and plot in Fig.\ref{fig:resolved-dephasing} the contribution
of the Fock state $|n\rangle$ to the linewidth, that can be read off 
from the operators under the trace in 
Eqs.(\ref{eq:result-linewidth},\ref{eq:McGowan-linewidth}).
In the same plot, we also give the photon statistics (peaked curves).
The chosen parameters correspond to the first trapping transition
just above $\theta \approx 2\pi$ where the maximum of the photon 
statistics jumps from lower to higher values. We see that 
Eqs.(\ref{eq:result-linewidth}) predicts a weak oscillation of the 
linewidth as the mean photon number increases, while 
Eq.(\ref{eq:McGowan-linewidth}) is essentially flat in this regime. 
These oscillations arise from the different prefactors $\hat\varphi$ 
and $\hat n^{1/2}$ of the
sine terms in Eq.(\ref{eq:result-linewidth}). Indeed, performing
and expansion in $1/\hat n$, the trigonometric functions 
can be simplified and give an expression identical in leading order
to Eq.(\ref{eq:McGowan-linewidth}). The remainder is of relative over
$1/\hat n^2$ and contains the sinusoidal oscillation seen in
Fig.\ref{fig:resolved-dephasing}.

% 
% two interesting cases: at threshold;
% near first trapping state.

% !! Attention, with $n_{\rm th} = 0$, one is just missing the trapping
% states ($\delta$-peaks). !!

% Plot: photon statistics and summand of linewidth in one plot, to 
% highlight `regions' of photon numbers that contribute particularly to 
% the linewidth. Problem: with zero thermal photons, very sharp 
% features around the trapping states. 

% In the `perfect' maser, the interaction $\tau$ is fixed and does not
% fluctuate. We consider here the limit where the sum over the Fock 
% states can be replaced by an integral. If the 
% photon statistics is single-peaked with a sufficiently large width 
% (many photons), one gets
% \ldots
% which is identical to Eq.(10) of the classical Ref.\cite{Scully91}.

\begin{figure}[bth]
    \centerline{\includegraphics*[width=70mm]{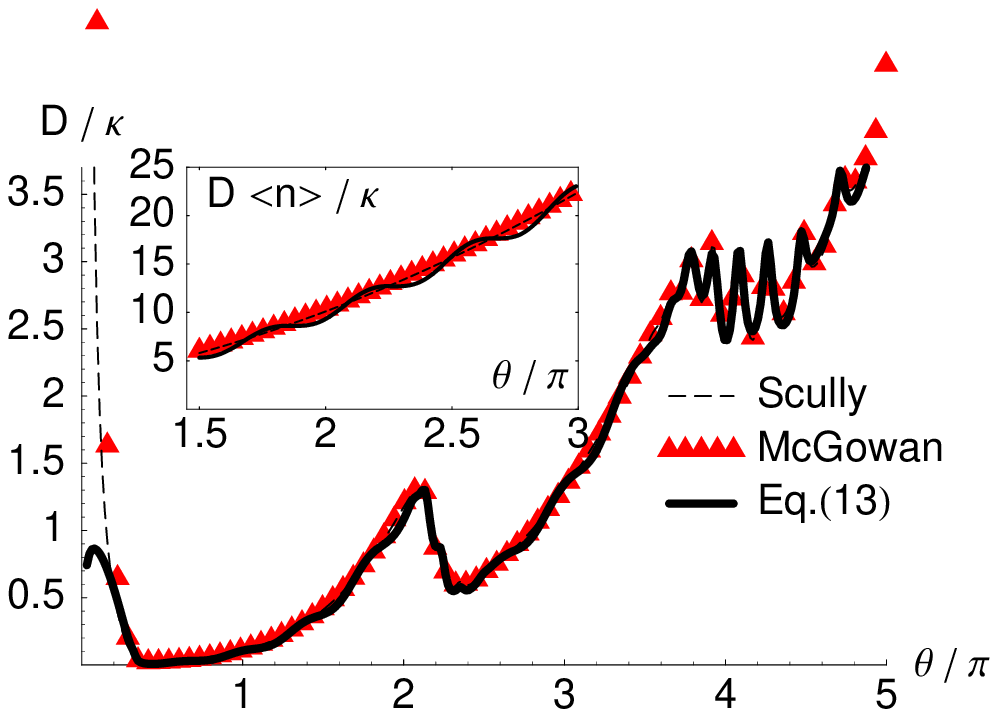}}
    \caption[]{Comparing expressions for the micromaser linewidth
    vs.\ the pumping parameter $\theta = (r/\kappa)^{1/2} g \tau$. No 
    average over the interaction time $\tau$ here. Main plot: the expressions 
    by Scully [Eq.(\ref{eq:Scully-linewidth}), dashed line] and by
    McGowan and Schieve [Eq.(\ref{eq:McGowan-linewidth}), red triangles] 
    practically coincide, except far below threshold ($\theta/\pi \ll 
    1/\pi$). Eq.(\ref{eq:result-linewidth}) suggested here predicts in 
    addition weak oscillations. 
    Inset: linewidth normalized to the Schawlow-Townes value,
    $D \langle \hat n \rangle_{\rm ss} / \kappa$ vs.\ pumping parameter 
    $\theta$ around the first trapping state. A similar smooth 
    behaviour is found also at larger values of $\theta$.
    \\
    Parameters: $r = 50\,\kappa$, $n_{\rm th} = 0.01$. We scan 
    $\theta$ via the product $g \tau$.}
    \label{fig:width}
\end{figure}

In Fig.\ref{fig:width}, we plot the linewidth as a function 
of the pump parameter $\theta$ and see that all three expressions 
give very similar values for $D$. We also note (see inset) that 
the product $D \langle \hat n \rangle_{\rm ss} / \kappa$ that measures 
the deviation from the Schawlow-Townes limit of a conventional laser
is essentially a quadratic function of $\theta$, similar to the 
formula found in Ref.\cite{Scully91}:
\begin{equation}
    D \langle \hat n \rangle_{\rm ss} \approx \frac{ \kappa \theta^2 + 
    \kappa( 1 + 2 n_{\rm th} ) }{ 4 }
    \label{eq:Scully-Schawlow-Townes}
\end{equation}
It is interesting to note that the sharp features of $D$ near the
trapping states essentially arise from the corresponding oscillations
in the average photon number.

\subsection{Exponentially distributed interaction time}
\label{s:laser}

Our formula~(\ref{eq:result-linewidth}) can be averaged explicitly
over an exponentially distributed interaction time $\tau$. The
result takes the form
% Exponential statistics for $\tau$: can be simplified to 
(for 
simplicity, we make the replacement $g \bar \tau \mapsto \bar g$) 
\begin{equation}
    D \langle \hat n \rangle_{\rm ss} = \kappa \,{\rm tr}
    \frac{ 
    \left[ 1 + \bar g^2 ( 3\hat n + 2 ) + \bar g^4 ( \hat n + 1 ) (4
    \hat n + 1 ) \right] \rho_{\rm ss}
    }{
    [ 1 + 4 \bar g^2 (\hat n+1)] [ 1 + 2 \bar g^2 (2\hat n+1) + \bar g^4 ] }
    \label{eq:exp-model-linewidth}
\end{equation}
The corresponding photon statistics is given by the recurrence relation
\begin{equation}
p_{n+1} 
= \frac{ 2 \bar\theta^2 }{ 1 + 4 \bar g^2 (n+1) }
p_{n}
\label{eq:photon-statistics-Sargent-Scully}
\end{equation}
where $\bar\theta = g \bar\tau ( r / \kappa )^{1/2}$ 
is the pumping parameter corresponding to $\bar\tau$. Here, we have
neglected the thermal effects and set $n_{\rm th} = 0$.
% and we have introduced
% for later use the rate $A = 2 r (g \bar\tau)^2 = 2 \kappa \bar\theta^2$ 
% that represents the linear gain.

The linewidth~(\ref{eq:exp-model-linewidth}) is plotted in 
Fig.\ref{fig:exp-model-linewidth} as a function of the (average) 
pump parameter $\bar\theta$. The sharp features at the trapping 
transitions are smoothed due to the averaging over $\tau$. Above the 
threshold, $\theta \ge 1/\sqrt{2}$, the linewidth is quadratic in 
$\theta$. This can be derived from the Scully 
formula~(\ref{eq:Scully-linewidth}) by assuming that the average 
photon number is large and expanding the sine to lowest order. 
This gives a term $r (g\tau)^2 / 4 \langle \hat n \rangle_{\rm ss} = 
\kappa \theta^2 / 4 \langle \hat n \rangle_{\rm ss}$, and
taking the average over an exponential distribution for $\tau$, we get
\begin{equation}
    D \approx \frac{ 1 + 2 \bar\theta^2 }{ 4 \langle \hat n \rangle_{\rm 
    ss} }
    \label{eq:exp-Scully}
\end{equation}
This is plotted in dashed in Fig.\ref{fig:exp-model-linewidth} and 
gives good agreement way above threshold.

\subsection{Uniform approximation to the master equation}
\label{s:uniform}

\begin{figure}[tbh]
    \centerline{\includegraphics*[width=70mm]{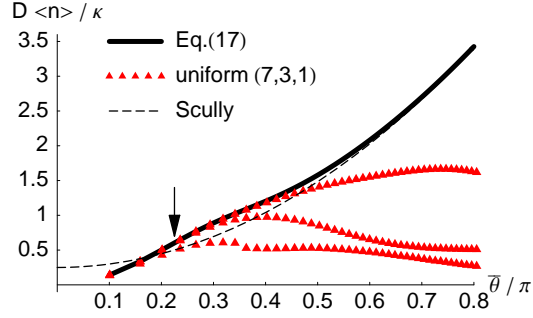}}
    \caption[]{Linewidth for a laser with exponentially distributed 
    coupling parameter $g \tau$.  Solid line from
    Eq.(\ref{eq:exp-model-linewidth}).  Dashed line:
    Eq.(\ref{eq:exp-Scully}) from Scully and co-workers.  Red
    triangles: uniform approximation (see Ref.\cite{Henkel07a}) up to
    order $7, 3, 1$ (top to bottom) in $g \bar\tau \hat \varphi$ for the
    expansion of $\sin( g \tau \hat\varphi )$ (see 
    Sec.\ref{s:uniform}). For the cosine, the 
    maximum order is $6, 2, 0$, respectively.
    \\
    Parameters: $g \bar \tau = 0.3$ and $\bar\theta = g \bar \tau
    (r/\kappa)$ is scanned via the pumping rate $r$.  The arrow marks
    the laser threshold $\bar \theta = 1/\sqrt{2}$ where the 
    small-signal gain $2 r (g \bar \tau )^2$ equals the
    loss rate $\kappa$  (see Ref.\cite{Orszag}).}
    \label{fig:exp-model-linewidth}
\end{figure}

We also give in Fig.\ref{fig:exp-model-linewidth} the results from a 
recently developed approximation to the laser master equation that is 
built to be valid beyond the weak coupling regime $g \bar\tau \ll 1$,
see Ref.\cite{Henkel07a}. The basic idea is to replace the large 
set of Lindblad 
operators in
Eqs.(\ref{eq:def-Lindblad-start})--(\ref{eq:def-Lindblad}) 
by a smaller one, using an expansion adapted to the probability 
measure ${\rm d}p( \tau )$. More specifically, we represent 
the operators $\cos(g \tau \hat\varphi)$ and $\sin(g \tau \hat\varphi)$ 
as sums of normalized orthogonal polynomials in $\tau / \bar\tau$, 
with coefficients that are rational functions of $g \bar\tau 
\hat\varphi$. The expansion is truncated at polynomials of a given 
order and gives good agreement for the average photon number
and its variance. The agreement persists above the laser threshold
where a simple expansion in $g \bar\tau$ fails, as illustrated in 
Ref.\cite{Henkel07a}. Fig.\ref{fig:exp-model-linewidth} shows, 
however, that for the laser linewidth, polynomials of fairly high 
order (up to the seventh) have to be taken into account to describe 
the strong pumping regime $\theta \gg 1$. In fact, if we compare
the $n$-dependent linewidth (inside the trace of
Eq.(\ref{fig:exp-model-linewidth})) with its uniform expansion, we see
a good agreement only up to photon numbers where $g \bar\tau n^{1/2}$ 
is comparable to the maximum retained order. Whenever the photon 
statistics covers photon numbers beyond that value, the (truncated) 
uniform approximation breaks down. This can be translated into a 
upper limit for the pumping parameter that scales linearly with
the order---a behaviour that can be qualitatively seen in
Fig.\ref{fig:exp-model-linewidth}.

%?? compare to exact calculation at fairly high temperature?

% If we differentiate this expression for $t \searrow 0$,
% we get a formula where $D$ is retrieved
% directly from the master equation:
% 
% where
% the density matrix $\rho_{\rm eq}$ is diagonal in the number state 
% basis. Its matrix elements $p_{n} = \rho_{nn}$ satisfy the recurrence 
% relation

% illustrate convergence of uniform approximation introduced in 
% Ref.\cite{Henkel07a}. Recall idea of expanding Lindblad operators.

% \begin{eqnarray}
%     \frac{ {\rm d} P }{ {\rm d}t } & = & 
%     \kappa \left[
%     a P(t) a^\dag - {\textstyle\frac{ 1 }{ 2 }}
%     \left\{ P(t) a^\dag a +  a^\dag a P(t) \right\}
%     \right]
%     \label{eq:q-regression}
%     \\
%     && {} +
%     r \int\!{\rm d}p( \tau )
%     \left[
%     \cos( g \tau \hat \varphi ) P( t ) \cos( g \tau \hat \varphi ) 
%     +
%     (g \tau)^2 a^\dag \sinc( g \tau \hat \varphi ) P( t ) 
%     \sinc( g \tau \hat \varphi ) a
%     \right]
%     \nonmber
% \end{eqnarray}
% identical to master equation~(\ref{eq:master-eqn}) except that  `skew'
% operator $P$

\section{Conclusions}

We have discussed in this paper an alternative calculation of the 
linewidth of a laser, with emphasis on the micromaser. The expression 
we suggest can be immediately computed from the master equation, 
using the quantum regression theorem, see Eq.(\ref{eq:q-regression-1}).
The result emerges as a average over the photon number statistics, 
similar to previous work that we find good agreement with. 

In our expression to the linewidth, a genuine contribution from a 
`dephasing' mechanism is evident. It is due to the cosine terms in 
Eq.(\ref{eq:result-linewidth}) that can be traced back to those 
pumping events where the pumping atom is detected again in the excited 
state after crossing the cavity. These processes cannot contribute to 
the laser gain, this is why only the sine terms appear in the photon 
statistics [Eq.(\ref{eq:steady-state-recurrence})]. The combined 
atom+cavity system does get affected, however, by a signed probability 
amplitude.  This is very similar to non-destructive measurements of
the photon number using atoms in superposition states~\cite{Nogues99}.
It is the average over these signed probability amplitudes, since they 
depend on the photon number, that contributes to the maser linewidth.
Our formula for the linewidth nicely embodies this
mechanism via the differences $\cos( g \tau \sqrt{n+1} ) - \cos( g
\tau \sqrt{n} )$ that occur in Eq.(\ref{eq:result-linewidth}).
Even when the interaction time is broadly distributed, dephasing 
contributes a slight increase of the linewidth near the lasing 
threshold (Fig.\ref{fig:exp-model-linewidth}).

\smallskip

\paragraph{Acknowledgements.}
I thank Giovanna Morigi and the colleagues from the quantum optics 
workshop in Palermo (June 2007) for instructive discussions.

\appendix

\section*{Appendix}%Calculation of the linewidth}
\label{a:work-out-linewidth}

The contribution of the thermal absorption and emission terms 
to the linewidth is 
straightforward using the commutators
$\left[ a^\dag, L_{0} \right] = - \sqrt{ \kappa (1 + n_{\rm th})}$ and
$\left[ a, L_{1} \right] = \sqrt{ \kappa \,n_{\rm th}}$. This gives
\begin{equation}
    \left. D \right|_{\rm th}  \langle \hat n \rangle_{\rm ss}
    = 
    {\rm tr}\left[ \big(
    \kappa (1 + n_{\rm th}) \hat n - \kappa n_{\rm th} \hat n \big)
    \rho_{\rm ss} \right]
    \label{eq:loss}
\end{equation}
or $\left.  D \right|_{\rm th} = \kappa$.  (Perfectly reasonable: only
spontaneous, no stimulated processes contribute to the linewidth.)
The pumping terms can be simplified with the help of the identity
% (for simplicity,
% $g \tau_{k} \mapsto g$) 
\begin{equation}
    a^\dag F(\hat\varphi ) = 
    F(\hat n^{1/2} ) a^\dag,
    \label{eq:operator-trick}
\end{equation}
valid for any smooth function $F( \hat\varphi )$. For example, 
commutators are transformed into differences
\begin{equation}
    \left[ a^\dag, L_{{\rm c},k} \right] =    
    \left[
    \cos( g \tau_{k} \hat n^{1/2} ) - \cos( g \tau_{k} \hat \varphi )
    \right]
    a^\dag
    \label{eq:result-commutator}
\end{equation}
We thus get from the cosine terms
\begin{eqnarray}
    \left. D \right|_{\rm cos} 
 \langle \hat n \rangle_{\rm ss} &=& r \sum_{k}
\Delta p( \tau_{k} ) {\rm tr} \left\{
\left(
\cos( g \tau_{k} \hat \varphi ) - 
\cos( g \tau_{k} \hat n^{1/2} ) 
\right)^2
\hat n \rho_{\rm ss}
\right\}
\label{eq:result-linewidth-cos}
\end{eqnarray}
The result from the sine terms is slightly more complicated. 
We now use the detailed balance 
for the photon statistics to bring it into a manifestly positive form.
Recall that the steady state density matrix has diagonal elements
$p_{n} = \langle n | \rho_{\rm ss} | n \rangle$ that satisfy the
recurrence relation~\cite{Filipowicz86}
\begin{equation}
    p_{n+1} = \frac{ 1 }{ 1 + n_{\rm th} }
    \left( n_{\rm th} + 
    \theta^2 \sinc^2( g \tau \sqrt{ n + 1 } )  \right) 
    p_{n}
    \label{eq:steady-state-recurrence}
\end{equation}
Including the average over the interaction time $\tau$, we find
\begin{equation}
0 = - \kappa (1 + n_{\rm th} ) p_{n+1} +
\kappa n_{\rm th} p_{n} +
    \kappa \sum_{k} \Delta p( \tau_{k} )
    \theta_{k}^2 \sinc^2( g \tau_{k} \sqrt{ n + 1 } )  
    \,p_{n}
    \label{eq:steady-state-recurrence-1}
\end{equation}
with the pumping parameter $\theta_{k} = (r/\kappa)^{1/2} (g \tau_{k})$. 
We multiply this expression with $n+1$, sum over $n$ and subtract it 
from the linewidth.
There is partial cancellation of the terms proportional to $\kappa$ 
with the thermal contribution~(\ref{eq:loss}), and finally only the 
thermal occupation 
remains, as stated in Eq.(\ref{eq:result-linewidth}).

\end{document}